\documentclass[aps,preprintnumbers,nofootinbib,superscriptaddress,showpacs,letterpaper,twocolumn]{revtex4-1}
\pdfoutput=1
\usepackage{amsmath,amssymb,amsbsy}
\usepackage[utf8]{inputenc}
\usepackage{graphicx,booktabs,}
\usepackage{ulem,color,topcapt}
\usepackage[colorlinks=true, linkcolor=blue, filecolor=blue, urlcolor=blue, citecolor=blue, pdftex, plainpages=false]{hyperref}

\definecolor{orange}{rgb}{1.0, 0.5, 0}

\newcommand{\gGF}{\ensuremath{g_{\rm GF}^2} }

\newcommand{\MSbar}{\ensuremath{\overline{\textrm{MS}} } }

\newcommand{\vev}[1]{\ensuremath{\left\langle #1 \right\rangle} }

\newcommand{\fig}[1]{Fig.~\ref{#1}}

\begin{document}
\title{\texorpdfstring{Continuous renormalization group $\beta$ function from lattice simulations}{Continuous renormalization group beta function from lattice simulations}}
\author{Anna~Hasenfratz}
\affiliation{Department of Physics, University of Colorado, Boulder, CO 80309}
\author{Oliver Witzel}
\affiliation{Department of Physics, University of Colorado, Boulder, CO 80309}

\date{\today}

\begin{abstract}
We present a real-space renormalization group transformation with continuous scale change to calculate the continuous renormalization group $\beta$ function in non-perturbative lattice simulations.
 Our method is motivated by the connection between Wilsonian renormalization group and the gradient flow transformation. It does not rely on the perturbative definition of the renormalized coupling and is also valid at non-perturbative fixed points.   Although our method requires an additional extrapolation  compared to  traditional step scaling calculations, it has several advantages which compensates for this extra step even when applied in the vicinity of the perturbative fixed point. We illustrate our approach by calculating the $\beta$ function of 2-flavor QCD and show that lattice predictions from individual lattice ensembles, even without the required continuum and finite volume extrapolations, can be very close to the result of the full analysis.  Thus our method provides a non-perturbative framework and intuitive understanding into the structure of strongly coupled systems, in addition to being complementary to existing lattice  determinations. 
\end{abstract}
\maketitle

\section*{Introduction}

The renormalization group (RG)  $\beta$ function encodes the energy dependence of the running coupling. While the $\beta$ function is scheme dependent,  the number of its zeros, corresponding to infrared  and ultraviolet  fixed points (IRFP, UVFP),  as well as the slope around the zeros are universal. The  characteristic structure of the $\beta$ function distinguishes conformal vs.~confining,  asymptotically vs.~infrared free, trivial vs.~asymptotically safe  systems \cite{Gross:1973ju,Politzer:1973fx,Caswell:1974gg,Banks:1981nn,Baikov:2016tgj,Shrock:2013cca,Antipin:2017ebo,Gorbenko:2018ncu}.  The perturbative $\beta$ function of 4-dimensional non-abelian gauge-fermion systems are known up to 5-loop level in the \MSbar scheme, but the perturbative expansion is unreliable at strong couplings \cite{Baikov:2016tgj,Ryttov:2016ner,Ryttov:2010iz,Ryttov:2016hal}.
 The  $\beta$ function calculated  non-perturbatively is essential to describe strongly coupled systems whether QCD-like, within the conformal window, or infrared free.

 The gradient flow (GF) renormalized coupling~\cite{Narayanan:2006rf,Luscher:2009eq,Luscher:2010iy}  is used in many  lattice calculations to study the non-perturbative  properties of strongly coupled systems~\cite{Fodor:2012td,Fritzsch:2013je,Fodor:2014cpa,Hasenfratz:2014rna,Fodor:2015baa,Hasenfratz:2016dou,Fodor:2015zna,DallaBrida:2016kgh,Chiu:2016uui,Chiu:2017kza,Fodor:2017gtj,Hasenfratz:2017qyr,Fodor:2017die,Chiu:2018edw,Leino:2019qwk,Hasenfratz:2019dpr,Witzel:2019jbe}. 
Most lattice studies consider  the discrete step-scaling function, where the  GF time $t$  and corresponding energy scale $\mu=1/\sqrt{8t}$  are tied to the volume $t = (cL)^2/8$~\cite{Fodor:2012td,Fritzsch:2013je,Nogradi:2016qek,Fodor:2014cpa}. The definition of the gradient flow renormalized coupling and the steps needed to take the continuum limit are justified perturbatively.
In this letter we rely on an alternative approach based on Wilsonian real-space renormalization group that is valid also at non-perturbative FPs \cite{Wilson:1971bg,Wilson:1971dh,Wilson:1973jj}. We develop a new method to predict the continuous $\beta$ function of the renormalized GF coupling. This method has several advantages compared to a traditional step-scaling calculation even when applied in the basin of attraction of the GFP.  This easily compensates for the one additional extrapolation required. The approach is general and equally applicable for confining, conformal, or infrared free systems.

\begin{figure}[tb]
\centering
\includegraphics[width=\columnwidth]{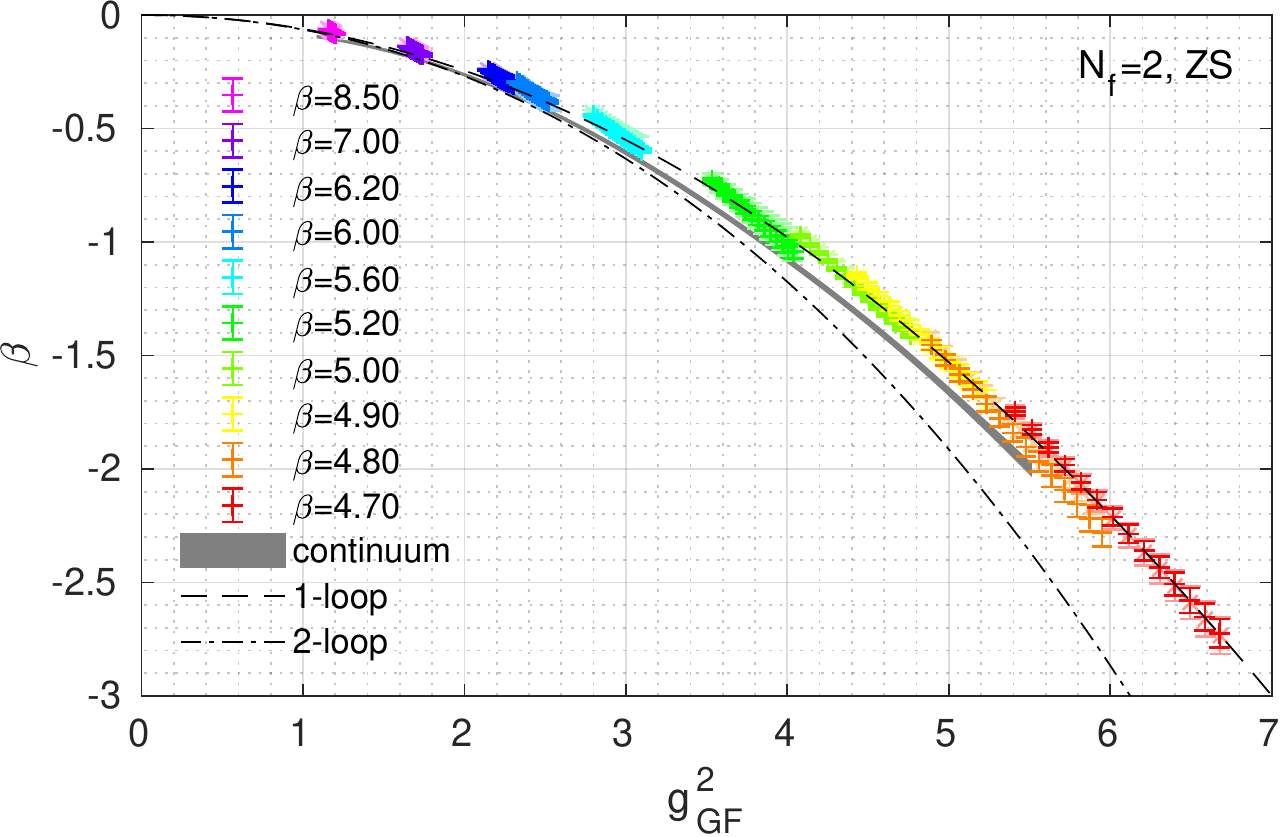}
\caption{Continuous RG $\beta$ function of 2-flavor QCD in the GF scheme. The grey band is the result of our full analysis with statistical uncertainties only. The colored data points show the lattice predictions for  $32^3\times64$ ('$+$') and $24^3\times64$ ('$\times$') ensembles in a wide  range of bare couplings without any extrapolation or interpolation.  Only flow times $t/a^2 \in(2.0,3.64)$ are shown. The dashed and dash-dotted lines are  the perturbative 1- and 2-loop $\beta(g^2)$ functions.}
\label{fig:beta-direct}
\end{figure}

We demonstrate the method by calculating the $\beta$ function of a QCD-like system with two flavors and SU(3) gauge group. The final prediction of this study with relatively low statistics  is shown by the grey band in Fig.~\ref{fig:beta-direct}. It is consistent with existing finite volume step scaling function calculations of 3-flavor QCD~\cite{DallaBrida:2016kgh} in that it is close to  the 1-loop perturbative prediction. More interesting is that the colored data points in Fig.~\ref{fig:beta-direct},  corresponding to raw lattice data  at finite volume, differ only slightly from the result of the full analysis. The continuous $\beta$ function approach predicts the running of the renormalized coupling in a transparent way where cut-off and finite volume effects are clearly  identifiable. This property could be particularly helpful when analyzing near-conformal systems or infrared free systems where the gauge coupling at the GFP is an irrelevant  parameter.  GF measurements of existing configurations of step-scaling studies can be reanalyzed to predict the continuous $\beta$ function without additional computational cost. Hence this method provides an alternative  to test systematical errors.

\begin{figure}[tb]
\centering
\vskip 0.75cm 
\includegraphics[width=\columnwidth]{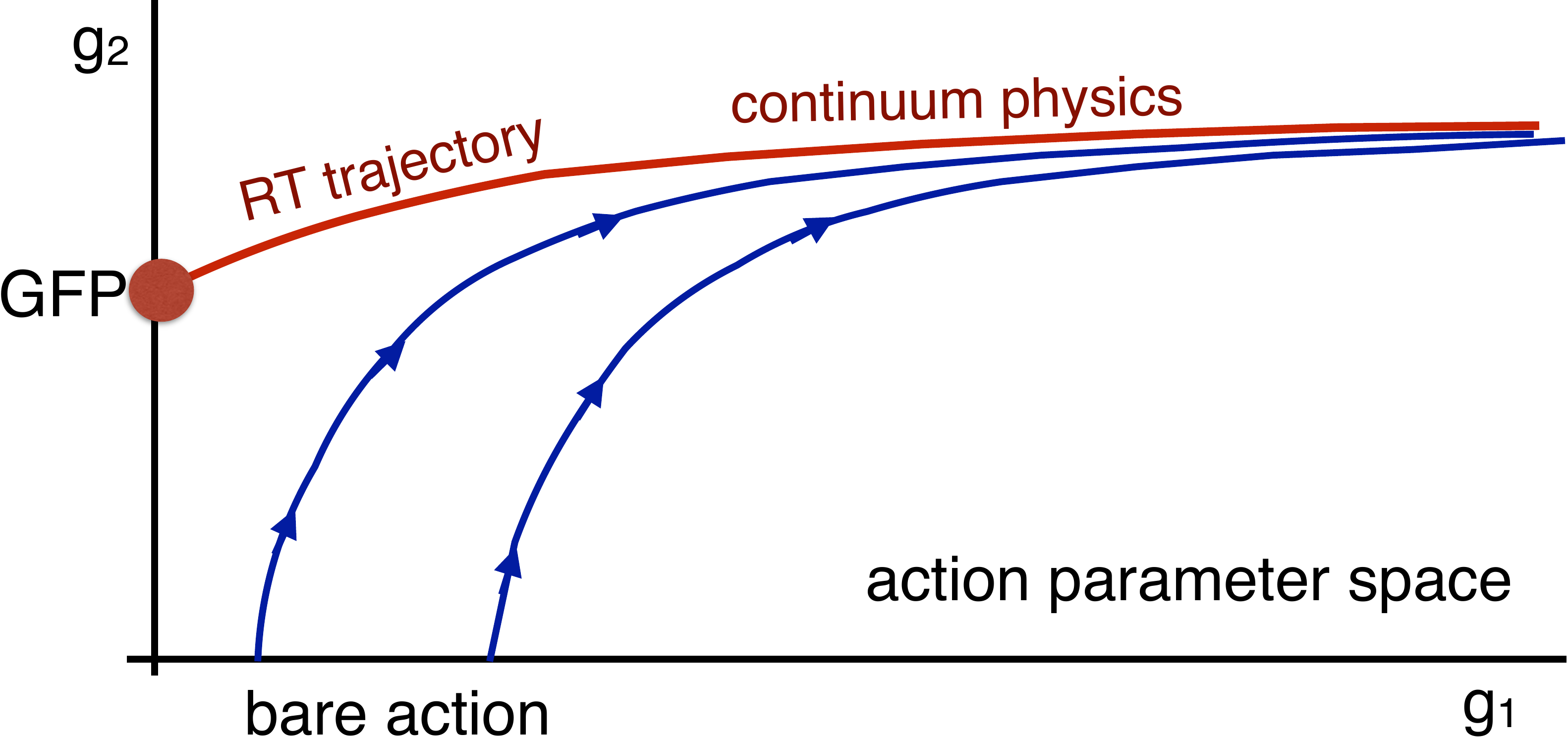}
\caption{Sketch of RG flow  in the multi-dimensional action parameter space.}
\label{fig:RG-flow}
\end{figure}

The connection between GF and RG is discussed  in Ref.~\cite{Carosso:2018bmz}.  Gradient flow is a continuous smearing transformation that is appropriate to define real-space RG blocked quantities, but it is not an RG transformation as it lacks the crucial step of coarse graining. However, coarse graining can be incorporated when calculating expectation values.  In particular, expectation values of local operators, like the energy density  that enters the definition of the GF coupling, are identical with or without coarse graining. When the dimensionless GF time $t/a^2$  is related to the RG scale change  as $b \propto \sqrt{t/a^2}$, the GF transformation  describes  a continuous real-space RG transformation. 

The  topology of  RG flow on the chiral $m=0$ critical surface in an asymptotically free gauge-fermion system is sketched  in \fig{fig:RG-flow}. $g_1$ represents the relevant gauge coupling at the Gaussian  fixed point (GFP), while $g_2$ refers to all other irrelevant couplings. The  GFP is on the $g_1=0$ (lattice spacing $a=0$) surface and the renormalized trajectory (RT) emerging from the GFP describes the cut-off independent continuum limit at finite renormalized coupling. 
Numerical simulations are performed with an action characterized by a set of bare couplings. If this action is in the vicinity of the GFP or its RT, the typical RG flow approaches the RT and follows it as the energy scale is decreased from the cut-off towards the infrared as indicated by the blue lines. RG flows starting at  different   bare couplings   approach the RT differently, but once the irrelevant couplings have died out, they all follow the same 1-dimensional renormalized trajectory and describe the same continuum physics. The RT of chirally broken systems continues to $g_1 \to \infty$, while conformal systems have an IRFP on the RT that stops the flows from either direction. While the topology of the RG space is universal, the location of the fixed points and their corresponding RTs  depend on the RG transformation. 

The RT is a 1-dimensional line, therefore, a dimensionless (zero canonical and zero anomalous dimension) local operator with non-vanishing expectation value  can be used to define a running coupling along the RT. The simplest such quantity in gauge-fermion systems is the energy density multiplied by $b^4$ (or $t^2$) to compensate for its canonical dimension. This is indeed the quantity defined in Ref.~\cite{Luscher:2010iy} as the gradient flow coupling
   $\gGF(t; g^2_0) \propto  \vev{t^2 E(t)}$.
$E(t)$, the energy density at flow time $t$,  can be estimated through various local lattice operators like the plaquette or clover operators. At large flow time  irrelevant terms in the lattice definition of $E(t)$ die out. In that limit $\gGF$ approaches a continuum renormalized running coupling and its derivative is the RG $\beta$ function
 \begin{equation}
\beta(g_{GF}^2) = \mu^2 \frac{d g_{GF}^2}{d \mu^2} = -  t \frac{d g_{GF}^2}{d t}.
\label{eq:contbfn}
\end{equation}
 The above definition   is valid in infinite volume only. In  a box of finite length $L$ the RG equation contains the term $L  (dg^2_{GF} /dL)$, a difficult to estimate quantity.  In our approach we extrapolate  $L/a \to \infty$  at fixed $t/a^2$ which  also sets the renormalization scheme  $c=0$. 
The continuum limit of the $\beta$ function is obtained at fixed $\gGF$ while taking $t/a^2\to\infty$. In QCD-like systems this automatically forces the bare gauge coupling towards zero, the  critical surface of the GFP.

The Wilsonian RG description suggests that lattice simulations at a single bare coupling can predict, up to controllable cut-off corrections, a finite part of the RG $\beta$ function. In  practice the finite lattice volume  limits the range where the infinite volume $\beta$ function is  well approximated. Chaining together several bare coupling values, we can cover the entire RT while the  overlap and deviation  between different volume and bare coupling predictions characterizes the finite volume and finite cut-off effects as illustrated in Fig. \ref{fig:beta-direct}. 

Once the GF coupling is determined and its derivative is calculated as the function of the GF time, the continuous $\beta$ function calculation requires two steps:
\begin{itemize}
\item[A)]  Infinite volume extrapolation at every GF time. 
\item[B)] Infinite flow time extrapolation at every $g^2_{GF}$.
\end{itemize}
Step B) removes cut-off effects and replaces the $L/a \to \infty$ continuum limit extrapolation of the step-scaling function approach. Step A)  is  new in the continuous $\beta$ function approach but is compensated by several advantages. In all GF analysis the flow time has to be chosen large enough to remove all but the largest irrelevant operator even on the smallest volume considered. In traditional step-scaling calculations the flow time grows with $L^2$ which leads to large statistical errors on the largest volumes.
In the continuous $\beta$ function approach the flow time is independent of the volume. This significantly reduces  statistical errors.  Finally the continuum limit of the continuous $\beta$ function  is obtained by extrapolating a continuous function of the flow time.  Although the data  are highly correlated, a continuous function nevertheless allows  control  to determine the functional form,  e.g.~the scaling exponent of the irrelevant operator. The correlations themselves can be handled by a fully correlated analysis similar to fitting subsequent time slices of a 2-point correlation function.

The RG $\beta$ function is defined in the chiral limit. Finite fermion mass  introduces a relevant operator that drives the RG flow away from the critical surface. Thus simulations with $am=0$ are essential to avoid yet another $am\to 0$ extrapolation.  This is always possible in chirally symmetric  conformal systems and can be enforced in chirally broken models by limiting the simulation volumes to be finite in physical units, below the inverse of the critical temperature.   The same constraint is present in step-scaling studies. 

A continuous $\beta$ function based on GF around the GFP  has been considered before. The only published result is a prediction at one renormalized  gauge coupling \cite{Fodor:2017die} that assumes perturbative scaling.  

\section*{Simulation details}

Our lattice study of 2-flavor QCD is based on tree-level improved Symanzik gauge action and chirally symmetric M\"obius domain wall (DW) fermions with stout smeared gauge links. Using \texttt{Grid} \cite{Boyle:2015tjk,GRID} we generate configurations at 10  bare gauge couplings $\beta_0=6/g_0^2$  ranging from 4.70 to 8.50 on  $16^3\times64$, $24^3\times 64$, and $32^3\times 64$ volumes with fermion boundary conditions periodic in space, antiperiodic in time. In this pilot study each ensemble has between 90 to 200 measurements separated by 10 molecular dynamics time units (MDTU). The simulations are performed setting the input quark mass to zero and  the 5th dimension of DW fermions to $L_s=12$. This leads to residual masses $a m_\text{res} < 10^{-6}$ for all gauge couplings considered.  The same combination of actions has been used in recent works \cite{Hasenfratz:2017qyr,Hasenfratz:2019dpr,Witzel:2018gxm,Witzel:2019oej} and properties for QCD simulations are reported in \cite{Kaneko:2013jla,Hashimoto:2014gta,Noaki:2014ura,Tomiya:2016jwr}.  Measurements for three different gradient flows, Wilson (W), Symanzik (S), and Zeuthen (Z), have been implemented in \texttt{Qlua} \cite{Pochinsky:2008zz,qlua} and we analyze three operators, Wilson plaquette (W), clover (C),  and Symanzik (S) to estimate the energy density \cite{Ramos:2014kka,Ramos:2015baa}.  Our data analysis is performed using the $\Gamma$-method which is designed to estimate and account for autocorrelations \cite{Wolff:2003sm}. 

\section*{Steps of the $\beta$ function calculation}
In this Section we demonstrate the calculation of the continuous $\beta$ function step by step. Additional information including a preliminary analysis of the SU(3) system with 12 fundamental flavors can be found in Ref.~\cite{Hasenfratz:2019puu}.

The GF coupling is defined as
\begin{align}
    \label{eq:pert_g2}
    \gGF(t; L,g^2_0) = \frac{128\pi^2}{3(N^2 - 1)} \frac{1}{1+\delta(t/L^2)} \vev{t^2 E(t)}\,.
\end{align}
The normalization  ensures that $\gGF$ matches the $\MSbar$ coupling at tree level, and the term $1/(1+\delta)$  corrects for the gauge zero modes due to periodic gauge boundary conditions~\cite{Fodor:2012td}.  On $L^3\times L_t$ volumes  
\begin{align}
    \label{eq:delta}
    \delta = - \frac{ \pi^2 (8t)^2}{L^3 L_t}  + \vartheta^3 \left(  \frac{t}{L^2} \right) \vartheta\left(\frac{t}{L_t^2}\right)\, ,
\end{align}
where $\vartheta$ is the standard Jacobi elliptic function~\cite{Fodor:2012td}. We calculate   $\beta(\gGF) = - t \, d\gGF(t) / dt$ using a symmetric 4-point numerical approximation for the derivative.

\subsection{Infinite volume extrapolation}

 \begin{figure}[tb]
\centering
\includegraphics[width=\columnwidth]{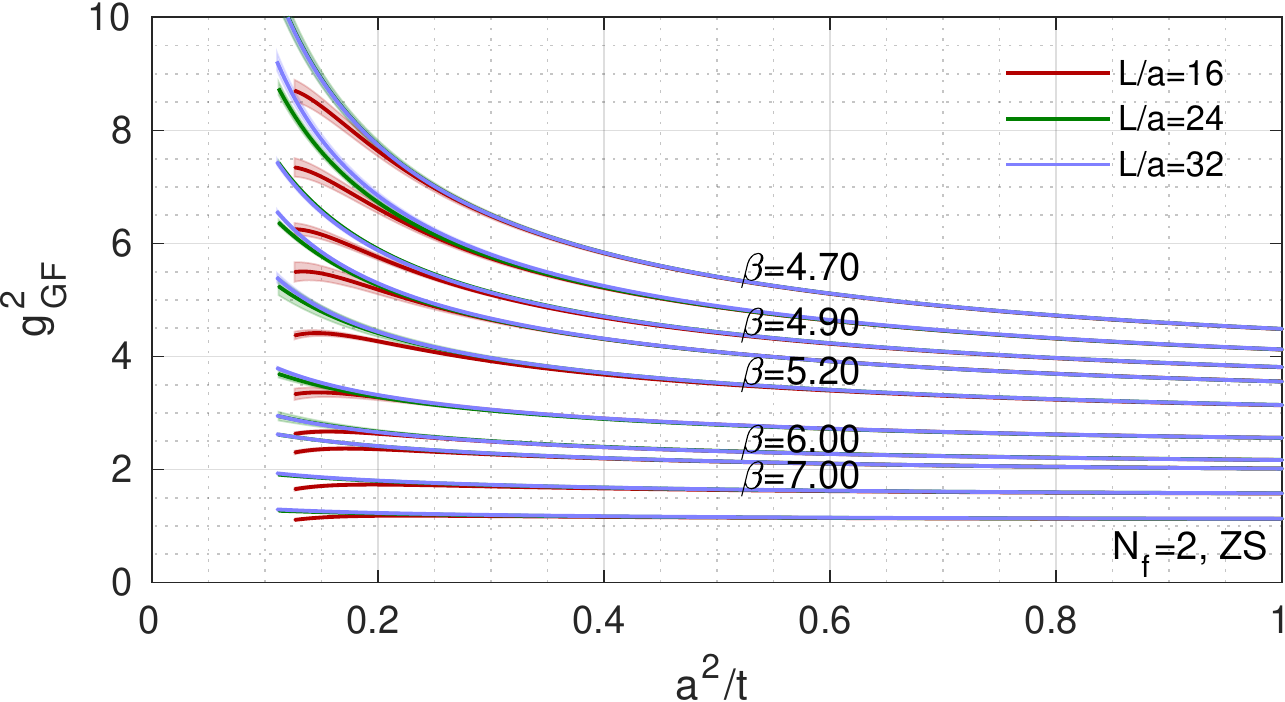} 
\caption{$\gGF$ as the function of $a^2/t$ at bare couplings $4.70\le \beta\le 8.50$ on the three different volumes we consider.}
\label{fig:g2_vs_t}
 \end{figure}
 
 \begin{figure}[tb]
\centering
\includegraphics[width=0.494\columnwidth]{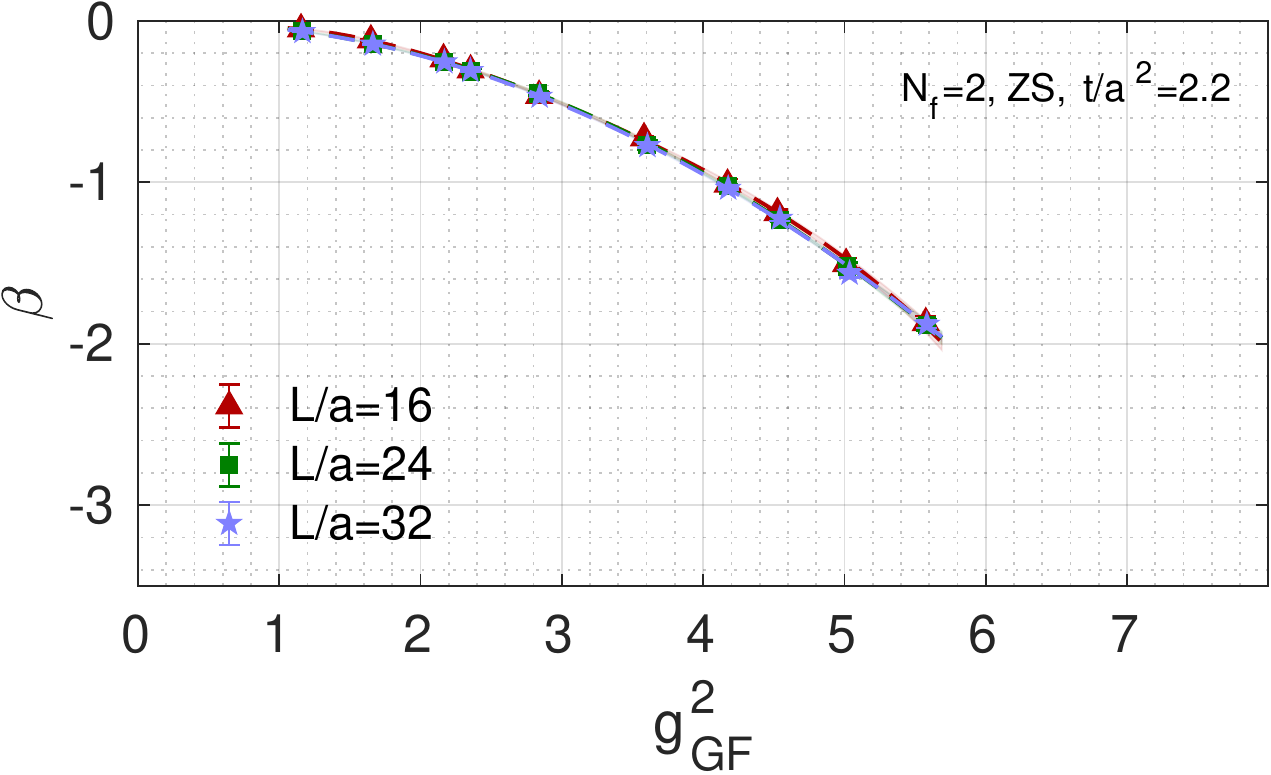}
\includegraphics[width=0.494\columnwidth]{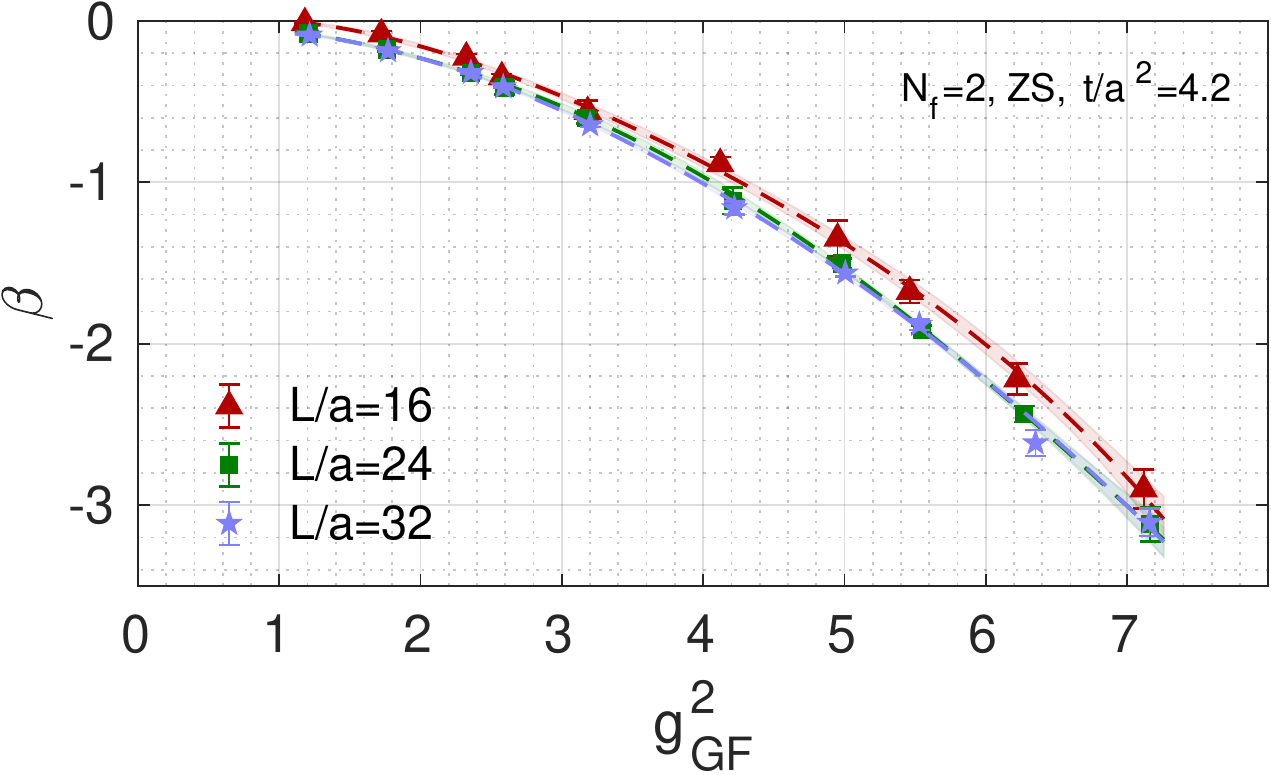}\\
\includegraphics[width=0.494\columnwidth]{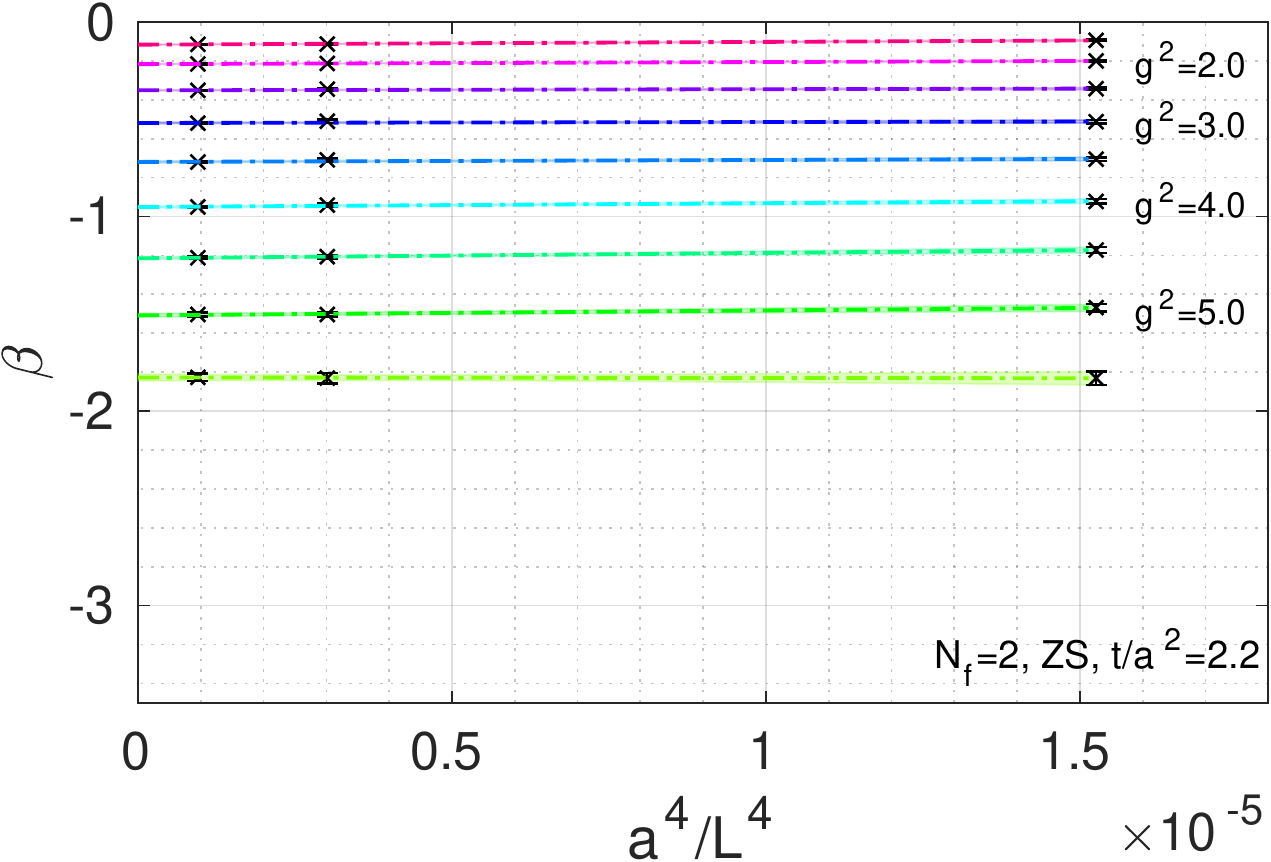} 
\includegraphics[width=0.494\columnwidth]{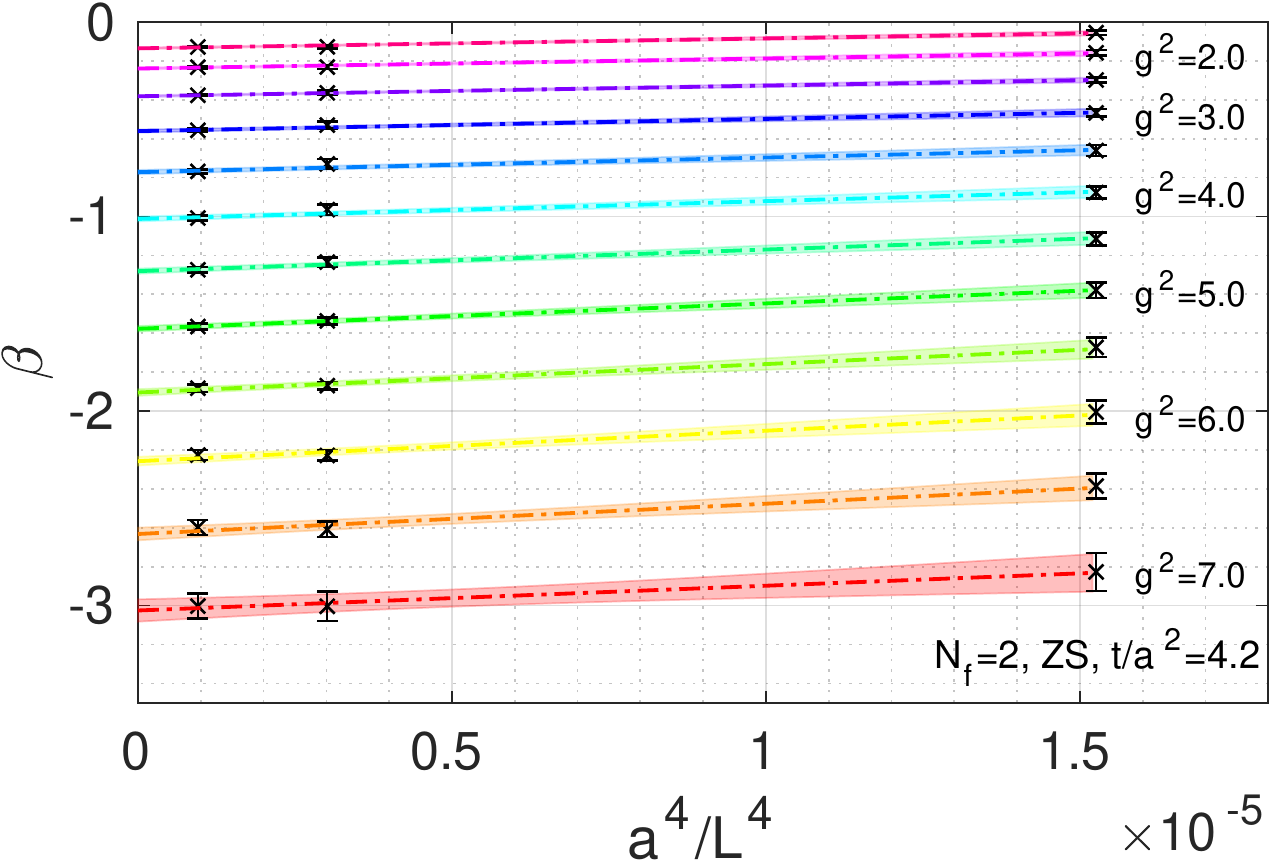}
\caption{ Top panels show the finite volume $\beta$ function at flow times $t/a^2=2.2$ and 4.2 ($a^2/t=0.455$ and $0.238$) for our three volumes. Dashed lines show a polynomial interpolation of the data points. Bottom panels present the infinite volume extrapolation at several $\gGF$ values for the same flow times.}
\label{fig:Linf_extrap}
\end{figure}

 The finite volume effects depend on $t/L^2$, and at leading order are proportional to $t^2/L^4$. Figure \ref{fig:g2_vs_t} shows $\gGF$ as the function of $a^2/t$ for our set of bare coupling values  on all three volumes for  Zeuthen gradient flow with Symanzik operator (ZS). The $L/a=16$ ensembles exhibit growing finite volume effects for $a^2/t\lesssim 0.2$, but the two larger volumes remain close up to $a^2/t\lesssim 0.1$.  We monitor both $\gGF$ and $\beta(\gGF)$ closely and  restrict the GF  time such that the finite volume corrections remain very small and the leading order  $t^2/L^4$ term is sufficient to take the infinite volume limit. 

 The $L/a\to\infty$ limit has to be taken at fixed  $t/a^2$ flow time  and coupling $g^2_{GF}$. We therefore determine  $\gGF(t)$ and its derivative on every ensemble and  interpolate $\beta(\gGF(t);L)$   with a 4th order polynomial for each lattice volume to predict the finite volume $\beta$ function at arbitrary renormalized coupling. The top  panels of Fig.~\ref{fig:Linf_extrap} illustrate this for the ZS combination.  We predict the infinite volume  $\beta(\gGF)$  using a linear $a^4/L^4$ extrapolation of the interpolated $\beta(\gGF)$ values. The lower panels of Fig.~\ref{fig:Linf_extrap} show  examples for $\gGF$ values spanning the accessible range in our numerical simulation. Finite volume effects are negligible at small flow time and the  extrapolations are mild, well described by a linear $a^4/L^4$ dependence even at $t/a^2=4.2$. As a consistency  check we compare extrapolations using all three volumes  to extrapolations using the two largest volumes only. While the errors of the infinite volume predictions change, the values are  consistent. Other flow and operator combinations show similar volume dependence.
 
\subsection{Infinite flow time extrapolation}
 \begin{figure}[tb]
\centering
\includegraphics[width=\columnwidth]{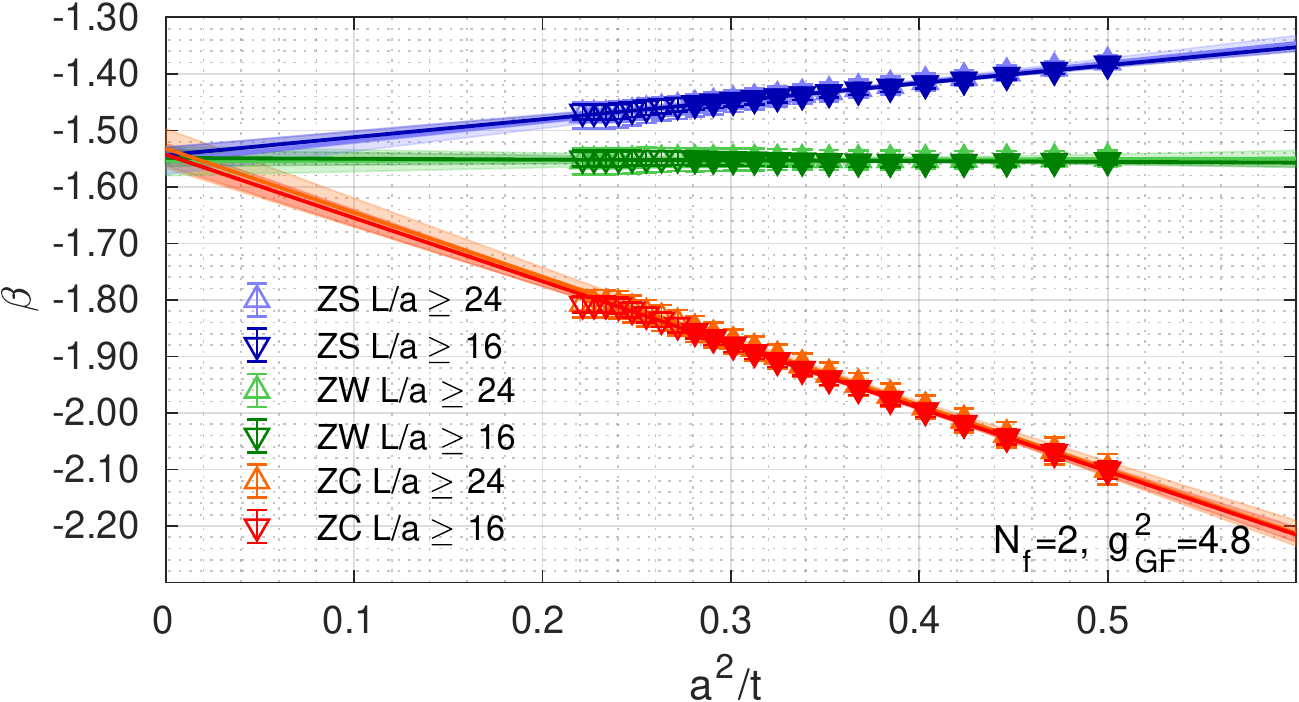}
\caption{Representative $a^2/t \to 0$ continuum limit extrapolation at $\gGF=4.8$. We show results for two infinite volume extrapolations and three operators, fitting filled symbols in the range $2.00 \le t/a^2 \le 3.64$ ($0.500\ge a^2/t\ge 0.274$). The (uncorrelated) fits are independent  but predict consistent $a^2/t=0$ continuum values.}
  \label{fig:cont_extrap}
\end{figure}
 The final step is the  $a^2/t \to 0$ continuum limit extrapolation at fixed $\gGF$. The GF time is a continuous variable but the range of $t/a^2$ values has to be chosen with care. First, the flow time must  be large enough for the RG flow to reach the RT where irrelevant operators are suppressed. Assuming one irrelevant operator  with scaling dimension $\alpha < 0$ dominates the cut-off effects, an extrapolation  in $(t/a^2)^{\alpha/2}$  predicts the continuum limit. Around the GFP $\alpha=-2$  and we find that our data is well described by a linear $a^2/t$ dependence when $t/a^2 \gtrsim 2.0$. Second, the upper end of the flow time range must also be restricted. When finite volume effects  depending on $t/L^2$ are large, the $L\to\infty$ extrapolations become unreliable. Figure \ref{fig:g2_vs_t} suggests that $t/a^2 \lesssim 4.0$ is sufficient to control this.  Any change of the continuum limit prediction when varying the minimal and maximal flow time values can be incorporated as systematical  uncertainty.   
 
 We show an example for the $a^2/t \to 0$ continuum extrapolation at $\gGF=4.8$ in Fig.~\ref{fig:cont_extrap} where we fit the data (filled symbols) in the range $2.00 \le t/a^2 \le 3.64$ ($0.500 \ge a^2/t \ge 0.274$).  In principle the flow time $t$ is a continuous variable; in practice we choose to dilute the data by fitting  in $\Delta t = 0.12$ intervals. Further we perform uncorrelated fits neglecting that values in $t$ are correlated which could easily be accounted for in a bootstrap or jackknife analysis.  Varying the minimal and maximal flow times in the range  $1.88 \le t/a^2 \le 4.06$ impacts the uncertainties but not the central values.

 We compare the  $\beta(\gGF;t)$ functions  obtained using Zeuthen flow and Wilson  plaquette, Symanzik, and clover operators in Fig.~\ref{fig:cont_extrap}.  We consider two different infinite volume extrapolations and  show for illustration additional data at larger flow time using open symbols.  The linear extrapolations in $a^2/t$ shown by the colored bands are obtained independently for each operator. Their  excellent agreement at the $a^2/t = 0$ limit is a  strong consistency check of the GF  time range and  the infinite volume extrapolation. 
Further consistency checks are possible by considering different flows. The scaling exponent $\alpha$ of the leading irrelevant operator could also be  extracted when performing  simultaneous fits to several operators.

\begin{figure}[tb]
\centering
\includegraphics[width=\columnwidth]{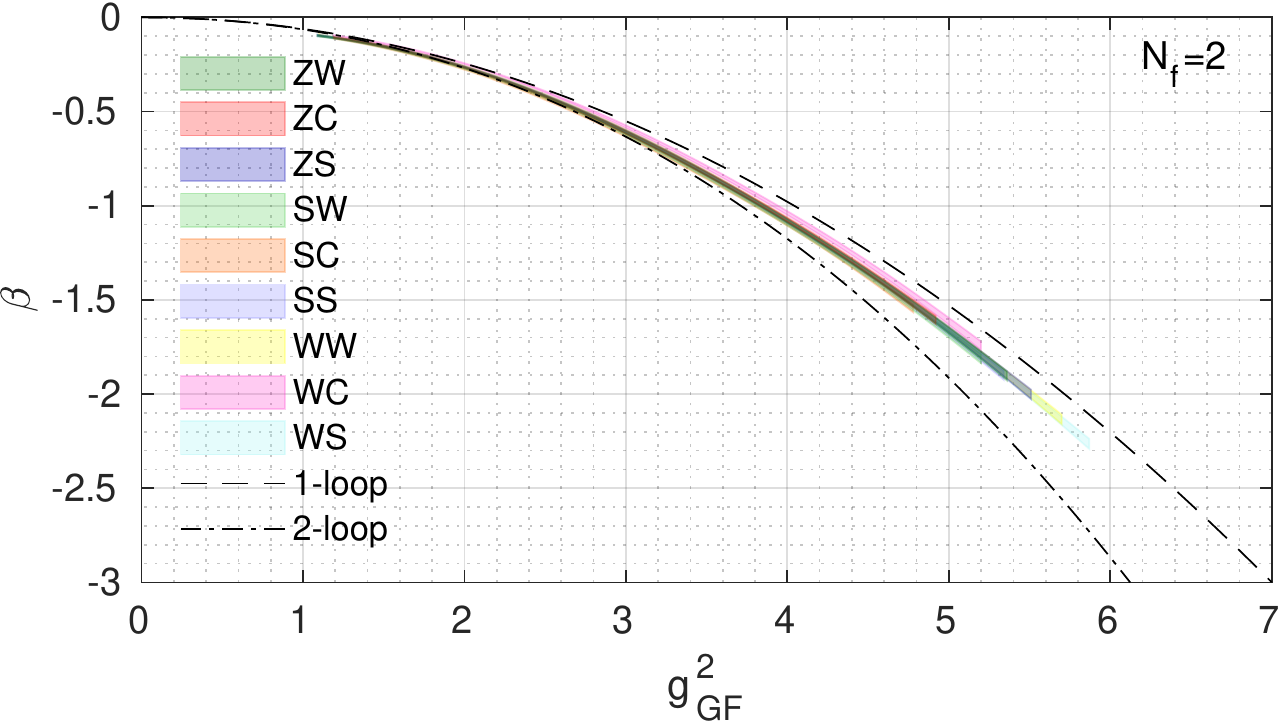}
\caption{Continuum limit of the continuous GF $\beta$ function predicted by nine different flow/operator combinations with statistical errors only. The different combinations are barely distinguishable and appear to be close to the 1-loop perturbative curve. }
\label{fig:beta-Zflow_continuum}
\end{figure}

\section*{Discussion}

We summarize the final result of our calculation by showing in Fig.~\ref{fig:beta-Zflow_continuum} the continuous GF $\beta$ function predicted from nine different flow and operator combinations. For reference we add the universal 1- and 2-loop perturbative predictions. The different flow/operator combinations are barely distinguishable, and the continuum limit prediction is very close to the raw ZS data as is shown in Fig.~\ref{fig:beta-direct}. The non-perturbative prediction follows the universal perturbative curves up to $\gGF\approx 2.0$, but at stronger couplings is closer to the 1-loop prediction. A similar trend is observed in Ref.~\cite{DallaBrida:2016kgh,Bruno:2017gxd} for 3-flavor QCD, suggesting that the GF coupling runs slower than the $\MSbar$ coupling. Since the RG $\beta$ function is scheme dependent, the GF and $\MSbar$ schemes do not have to agree. However the precise non-perturbative running is needed to determine $\alpha_s$, the $\Lambda_s$ parameter, or connecting lattice simulations to continuum results~\cite{Brida:2016flw,Bruno:2017gxd}.

The continuous $\beta$ function  described here works in both conformal or infrared free systems \cite{Hasenfratz:2019puu}. The relation between GF and Wilsonian RG is especially useful in strongly coupled conformal systems. The method complements the step scaling function studies, providing a new handle on systematic errors. In addition it provides an intuitive picture to understand the RG in numerical siumlations as we show in Fig.~\ref{fig:beta-direct}. 

\begin{acknowledgments}

We are very grateful to Peter Boyle, Guido Cossu, Anontin Portelli, and Azusa Yamaguchi who develop the \texttt{Grid} software library providing the basis of this work and who assisted us in installing and running \texttt{Grid} on different architectures and computing centers. We are indebted to Daniel Nogradi for extending his original calculation on symmetric volumes to asymmetric lattices and sharing the result prior to publication. We thank Alberto Ramos for many enlightening discussions during the ``37th International Symposium on Lattice Field Theory'', Wuhan, China, and Rainer Sommer and Stefan Sint for helpful  comments.  We benefited from many  discussions with Thomas DeGrand, Ethan Neil, David Schaich, and Benjamin Svetitsky. A.H.~and O.W.~acknowledge support by DOE grant DE-SC0010005.
A.H.~would like to acknowledge the Mainz Institute for Theoretical Physics (MITP) of the Cluster of Excellence PRISMA+ (Project ID 39083149) for enabling us to complete a portion of this work. 
O.W.~thanks the CERN theory group for their hospitality during the final stages of preparing this manuscript and acknowledges partial support by the Munich Institute for Astro- and Particle Physics (MIAPP) of the DFG cluster of excellence ``Origin and Structure of the Universe''. 

Computations for this work were carried out in part on facilities of the USQCD Collaboration, which are funded by the Office of Science of the U.S.~Department of Energy and the RMACC Summit supercomputer \cite{UCsummit}, which is supported by the National Science Foundation (awards ACI-1532235 and ACI-1532236), the University of Colorado Boulder, and Colorado State University. This work used the Extreme Science and Engineering Discovery Environment (XSEDE), which is supported by National Science Foundation grant number ACI-1548562 \cite{xsede} through allocation TG-PHY180005 on the XSEDE resource \texttt{stampede2}.  This research also used resources of the National Energy Research Scientific Computing Center (NERSC), a U.S. Department of Energy Office of Science User Facility operated under Contract No. DE-AC02-05CH11231.  We thank  Fermilab,  Jefferson Lab, NERSC, the University of Colorado Boulder, TACC, the NSF, and the U.S.~DOE for providing the facilities essential for the completion of this work. 

\end{acknowledgments}

\bibliography{../General/BSM}
\bibliographystyle{apsrev4-1} 
\end{document}